# Neural Code Summarization

**Piyush Shrivastava**
Khoury College of Computer Sciences
Shrivastava.pi@northeastern.edu

**Abstract**

Code summarization is the task of generating readable summaries that are semantically meaningful and can accurately describe the presumed task of a software. Program comprehension has become one of the most tedious tasks for knowledge transfer. As the codebase evolves over time, the description needs to be manually updated each time with the changes made. An automatic approach is proposed to infer such captions based on benchmarked and custom datasets with comparison between the original and generated results.

## Introduction

Code maintenance and improvement is a continuous cycle which along with changes in code requires providing a clear description of what the code is supposed to do and what changes were made. Maintaining updated code documentation, both on code and commit level can be a challenging task. The code segments follow a certain structure which if understood can reveal information about what it intends to accomplish. Moreover, a well-documented set of functions written for a different project can become handy to understand newly written code. With the abundance of open-source projects, leveraging the power of deep learning can yield a reduction of time spent on program comprehension.

The objective of here for the summary was to capture the structure of the function and to be able to generate summaries as code changes. The summaries are required to semantically accurate and reflect a developer's understanding of the code. The summary coherence is expected to change as the code structure changes. I use both preprocessed and programmatically annotated datasets for training and evaluation. Previous methods for code summarization have followed two main approaches: Heuristic based search and AI driven approach recently in practice. The report is based on application of more recent Sequence-to-Sequence models to transform text from one natural language domain to another. The task of code summarization matched the criteria for these techniques to be applied to simplify engineering practices and possibly automate code review to some extent.

## Requirements

- To explore the datasets used for training
- Creation of a custom dataset to capture the code diff granularity level in a git commit
- To be able to generate automated summaries
- Testing comprehension on different programming languages
- Evaluation based on BLEU and ROUGE scores
- Discuss human evaluation techniques

## Datasets

### Funcom

The Funcom dataset (LeClair et al., 2019) contains Java methods with 2.1 million lines of code in the form of function-comment pair along with function and comment ID. Both tokenized and filtered datasets were considered for data exploration and preprocessing. The function and comment ID can be used to group the individual segments with parent groups, from which a type of hierarchy can be generated. Hierarchical relation extraction was not considered for the dataset given the IDs, since the primary objective is to create summaries based on the code segments and as little metadata as possible. It can be observed from the generated histograms that function lengths are comparatively in the lower range.

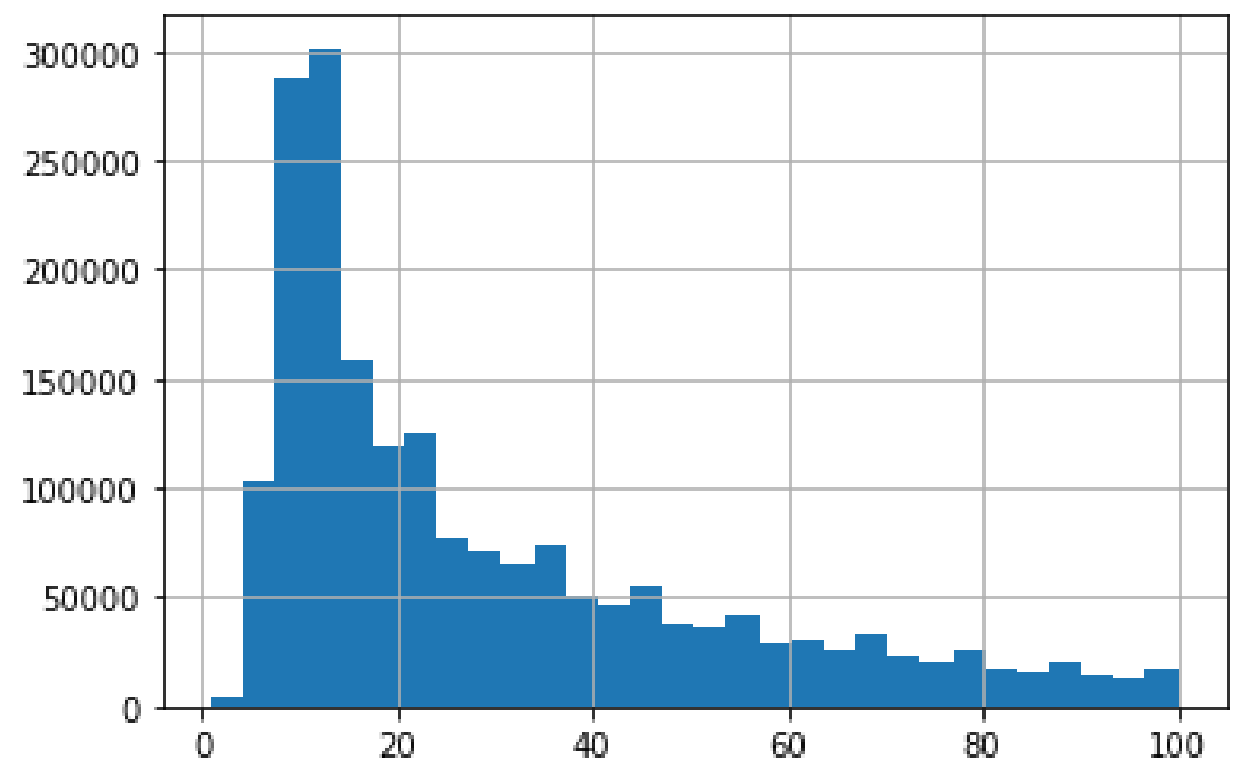

Frequency distribution for the Funcom tokenized dataset with function length

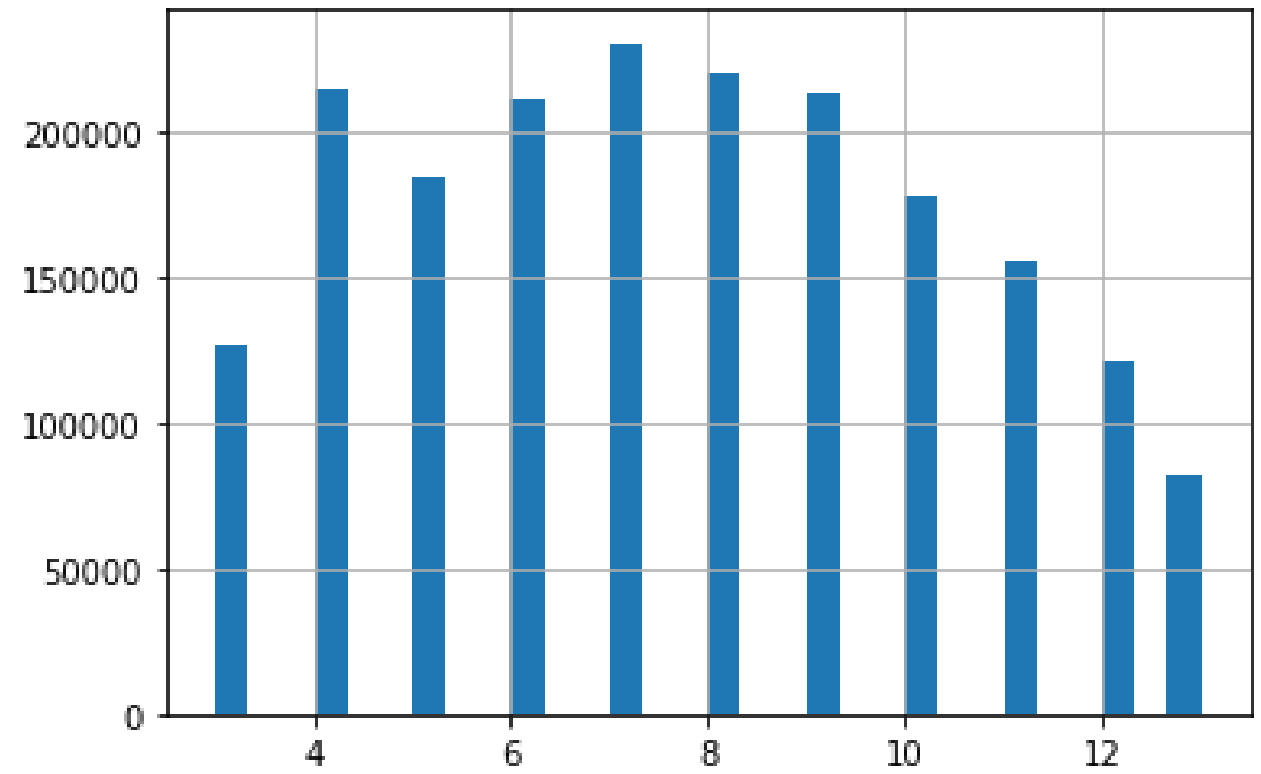

Frequency distribution for the Funcom tokenized dataset with comment length

The main difference between the tokenized and filtered datasets was the structure of functions, in which the latter had un-preprocessed functions and comments.

## Codesearch

Codesearch dataset (Husain et al., 2019) was available in languages such as Python, Java, Ruby and PHP. The primary focus of exploration was Python data. The dataset considered of code, docstring, code tokens and docstring tokens. The tokens were not part of the training process to keep the preprocessing like real-time data, where division into tokens can be difficult given custom variables and function declarations.

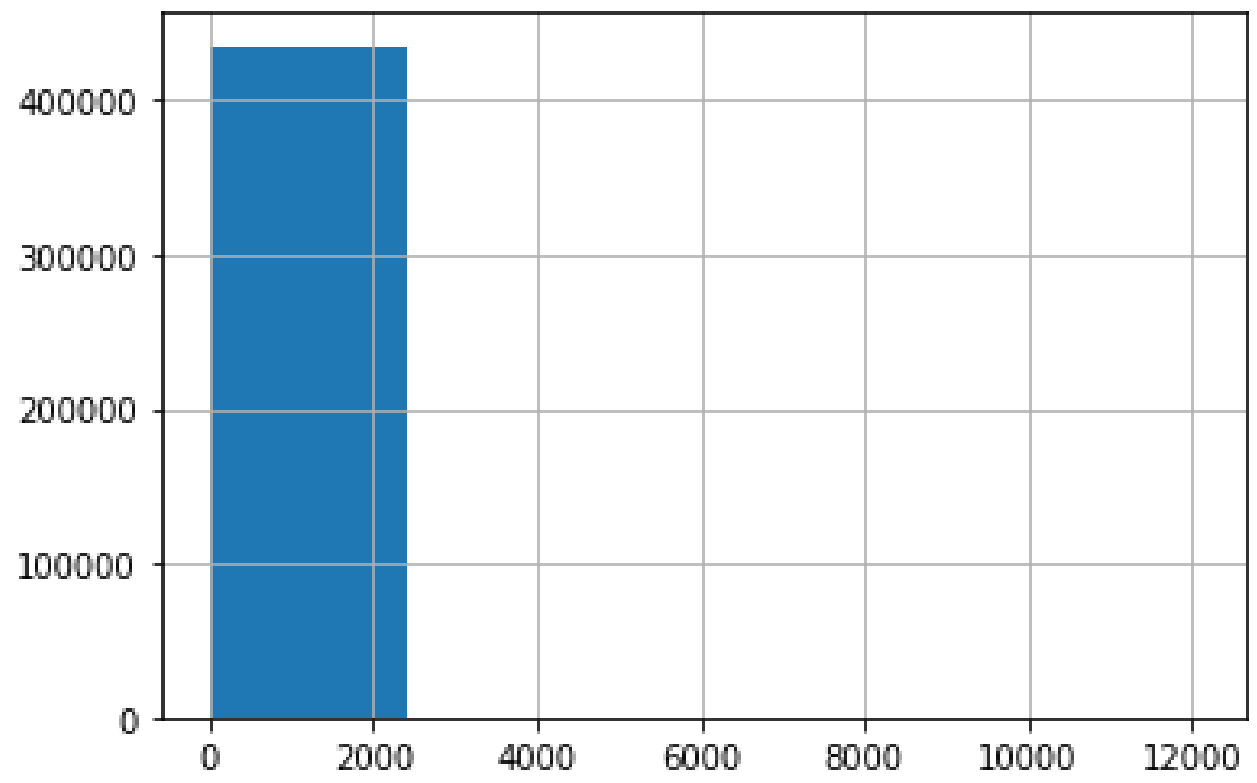

Frequency distribution for the Codesearch dataset with function length

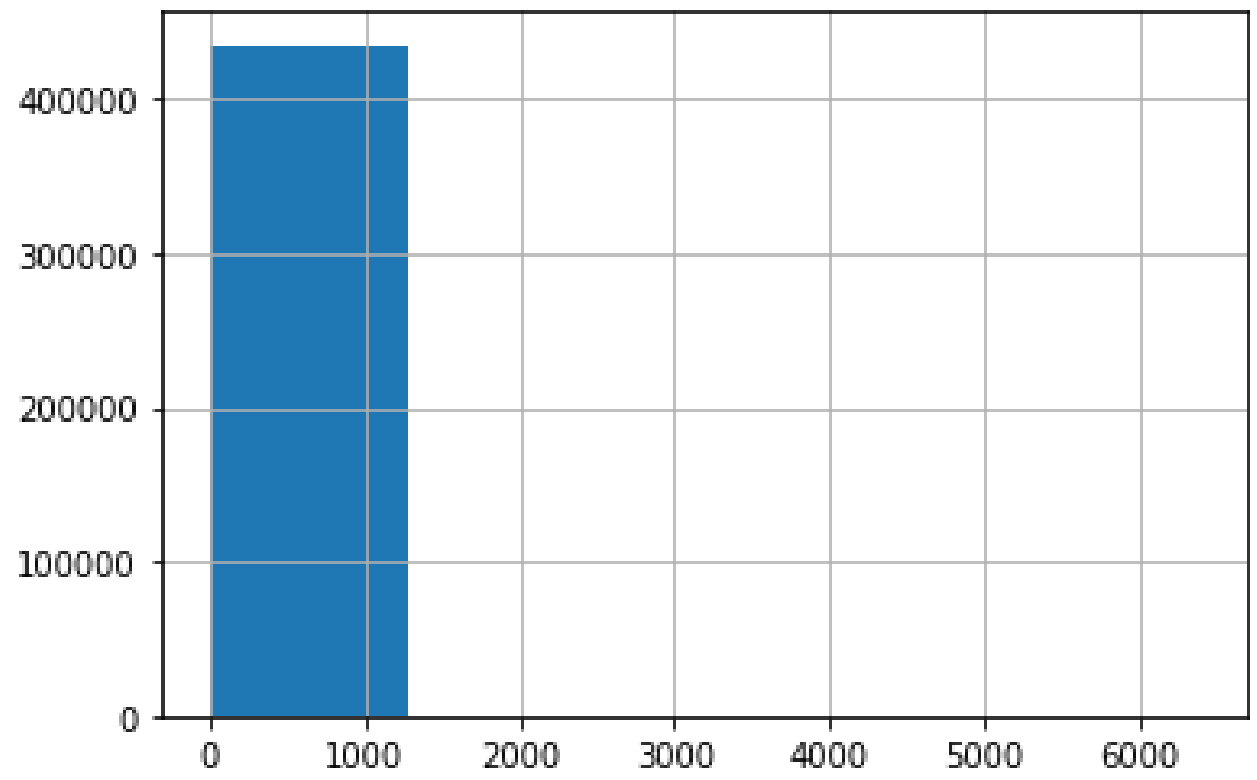

Frequency distribution for the Codesearch dataset with comment length

It can be observed the length is equally distributed between all types of functions and comments in Codesearch, as opposed to the varying distribution in Funcom dataset.

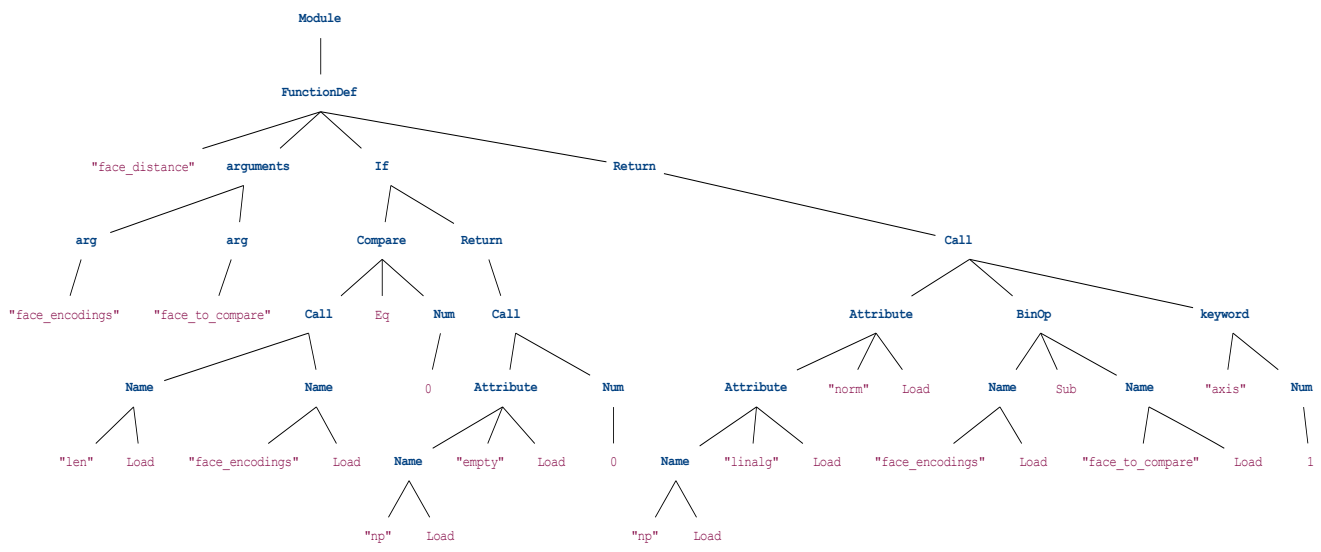

The syntax tree above generated by showast shows some of the system defined and custom named tokens in the set of functions.

### Pydriller-generated dataset

Pydriller is a Python library to obtain commit-related data from git sources. Consider a given repository and all the commits made in history. Then it is possible to obtain the commit message and corresponding source code differences between the current and last commit and provide annotations to create a custom dataset. But for each commit there can be multiple modifications, leading to multiple commit messages annotated as the same labels. Thus, it is important to either remove duplicates or consider adding a unique identifier such as name of functions in the modified file or timestamp. Popular C/C++ repositories such as boost and tensorflow were mined to gather commit message and diff insights with preprocessing applied to remove breaks and special characters added by GitHub to annotate a conflict. It is important to keep in mind that many repositories have a mixed distribution of programming languages used, so it can be expected in the pydriller dataset to occasionally encounter python or java tokens. As the commits contain code evolved over time, it is important to diversify the data before using it normalize the word count and avoid memorization by the model.

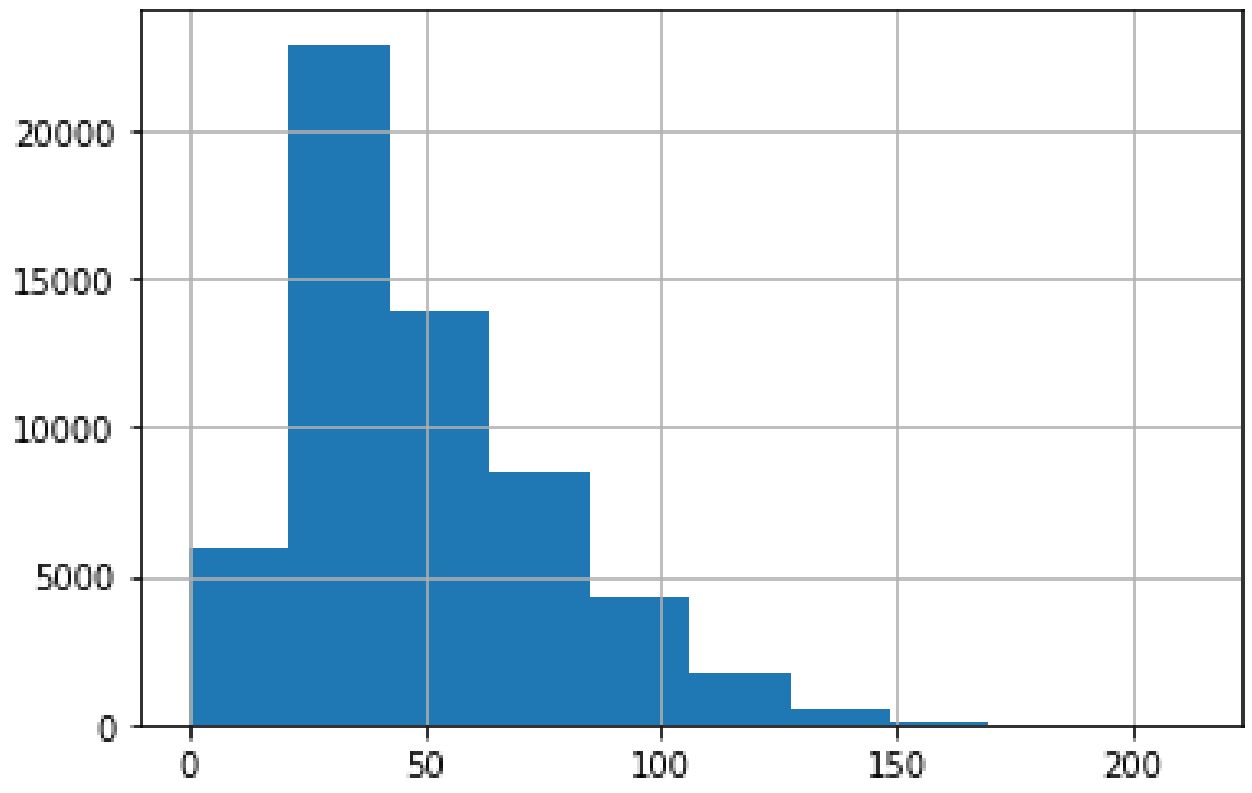

Pydriller dataset diff length frequency distribution

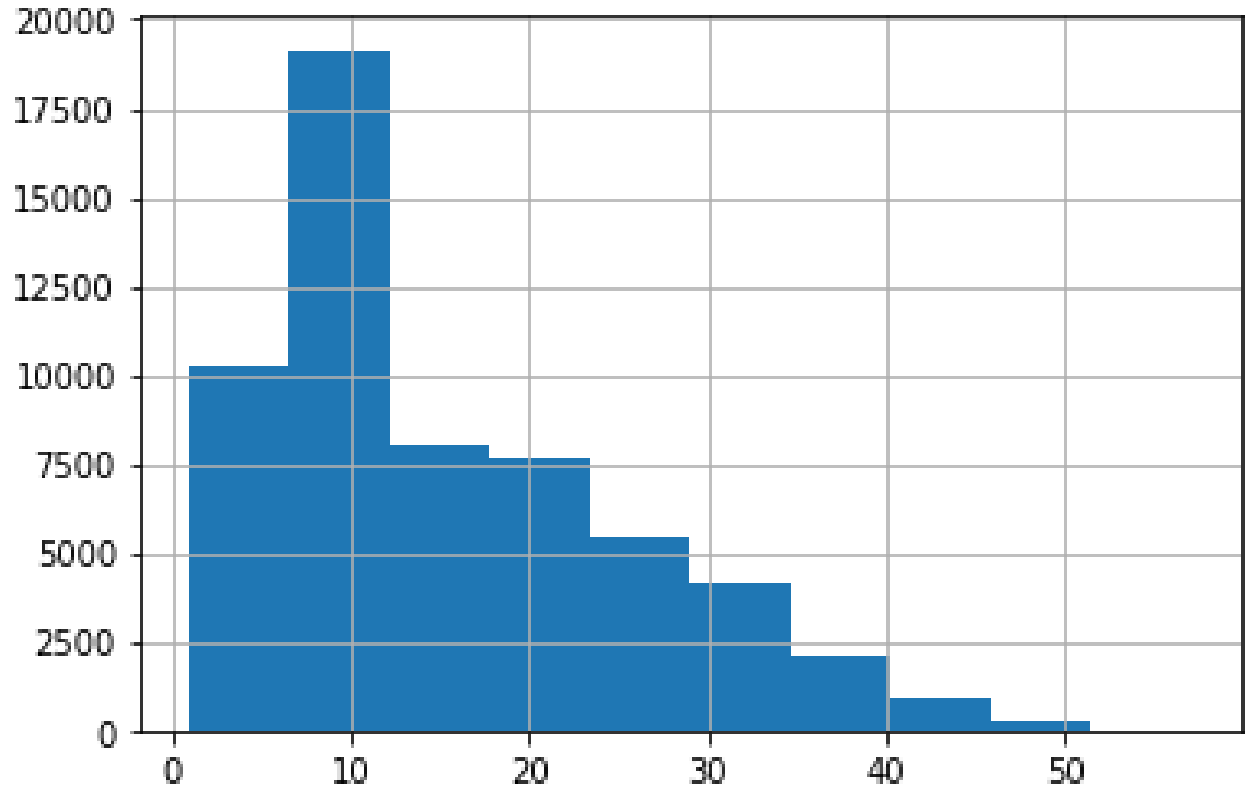

Pydriller dataset message length frequency distribution

## Problem

Data-driven source code summary generation involves application of NLP-based approaches to generate a description given a code segment as an input. The segment can be a few lines of code, a function, or a class object. For defining the problem, two levels of granularities are considered:

1. A code-comment pair
2. A commit message-diff pair

There are multiple approaches to proceed here in a supervised manner. The inputs can be defined as a stream of characters, generating a stream of transformed output characters. Alternatively, a stream of unigrams, bigrams or trigrams can be considered as the input source generated from the model. It would then be possible to pose this as a classification problem to predict a summary as a label. But given the amount of text in input here, the number of parameters in an n-gram model would be exceedingly large. The model is needed to generate a form of context which can help predict the output, and thus decode the sequence. Recurrent neural networks better fit the challenge, identifying it as a sequence-to-sequence transformation problem.

### Sequence-to-Sequence Models

Sequence-to-Sequence models (Sutskever et al., 2014) are based on an encoder-decoder architecture where both are separate neural networks combined into a model. The encoder network generates a representation of the input in a smaller dimension, while the decoder then takes this output as the input and generates an output representation.

Encoder-decoder example [Source: keras.io]

The tokenization of input generates a sequence of integers based on the model vocabulary to be padded for network to

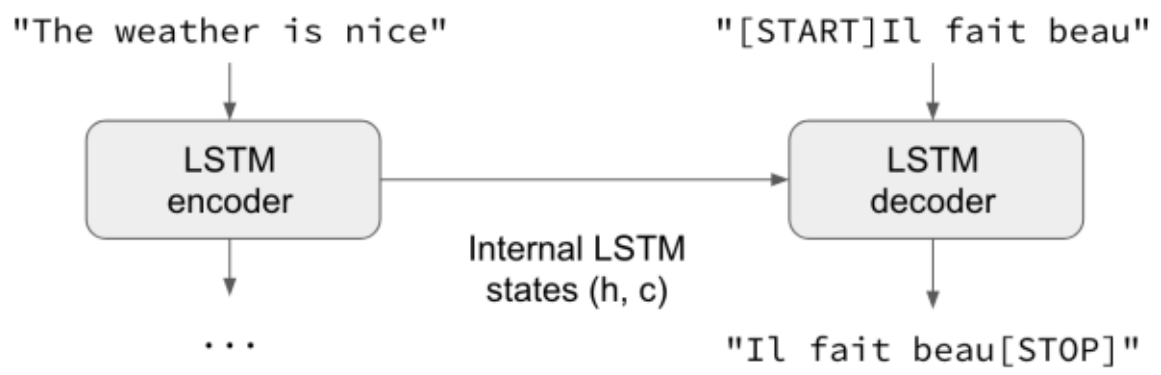

Encoder-decoder architecture [Source: blog.keras.io]

perform calculations. The hidden states and encoder outputs fed into the decoder produce an output distribution,

whereafter the input sequence can be decoded by sampling the token indexes one by one.

## Sequence-to-Sequence Models with Attention

For the problem defined here, a one-to-one LSTM architecture (Sutskever et al., 2014) consisting of an encoder-decoder framework may not produce the desired results, considering the difference of size between the input and output sequences. It can be challenging for the encoder to compress the entire sentence, while the result being a single representation of the source. It is possible that some parts of the input are more useful than others, and thus more worthy of attention. With attention, the model can be taught to pay attention to specific token with a greater relevance. Both encoder and decoder states are fed into the attention function to generate a scalar score for understanding relevance of a token for a particular step. There are many model variants for attention usage, such as dot-product, Luong attention (Luong et al., 2015), or Bahdanau attention (Chorowski et al., 2015), which was applied as the attention layer here.

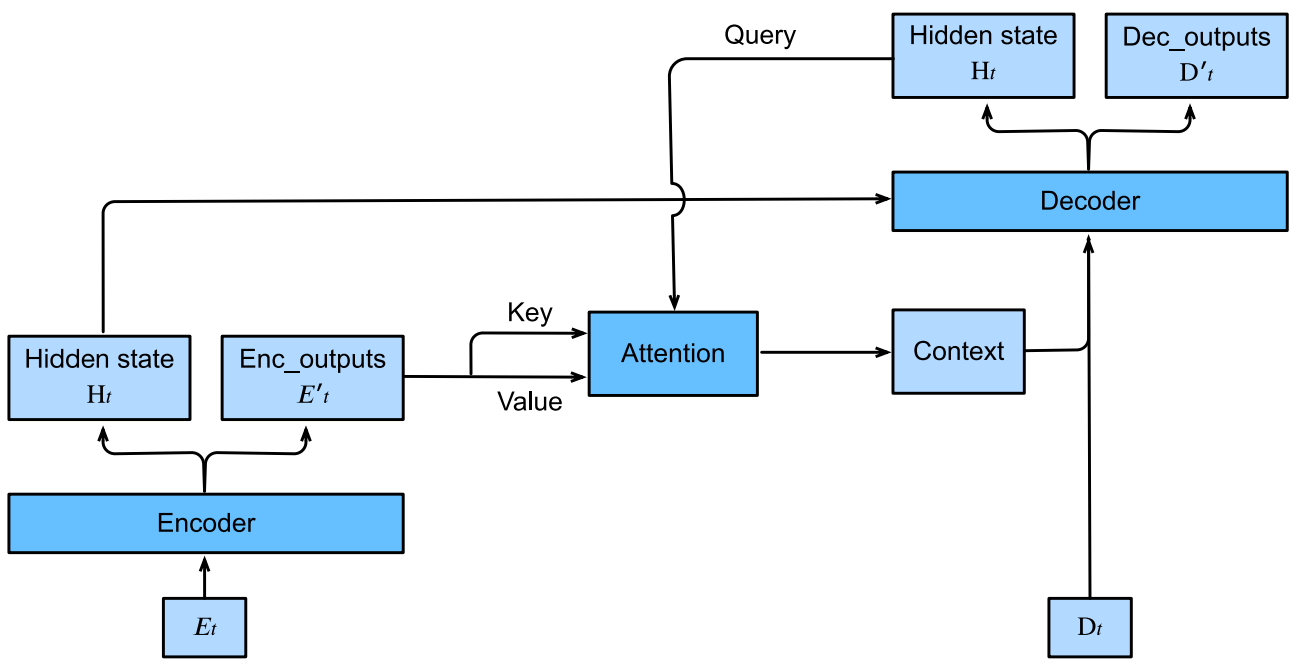

Seq2Seq attention mechanism [Source: d2l.ai]

# Experiments

One-to-one LSTM based sequence model (Sutskever et al., 2014) was experimented with initially starting with a sample of the Funcom dataset (LeClair et al., 2019). Tokenization was manual and at character-level. As expected, the model kept repeating the predicted sequences and often the characters within it. The same encoder-decoder architecture was tested with a shorter translation dataset to verify the implementation, and it seemed to provide more accurate results for the translation dataset, implying its unsuitability for the given problem.

The next step was experimentation with Bahdanau attention-based model (Chorowski et al., 2015) with keras tokenizer for generating the vocabulary for model. The encoder architecture was stacked LSTM with encoder embedding, with encoder outputs as initial state for the decoder LSTM along with its trainable embedding. This yielded more semantically accurate results to a relative degree with different sample sizes for Funcom and Codesearch datasets. Greedy decoding was used for generating the summaries by taking argmax over the probability distribution to sample the probable token index, with end tag for the decoder to understand the end of a sentence. The datasets were not experimented with to the full extent considering the GPU limitations involved.

With a greater amount of training data and a higher computational limit, the model is expected to provide a more comprehensive description to better describe the developer's understanding of code. Experimentation with the Pydriller dataset required tokenization with word limitation to a greater extent due to an increased percentage of rare words, which indicates the difference between a git commit message and a function comment, as well as git diff and the function itself. It is observed for Pydriller that model can learn to a greater extent if commits from a repository are not sequential and are not focused on a singular issue. The experiments and code are available on https://github.com/shrivastava-piyush/nlp-code-summarization.

# Evaluation

Use of evaluation metrics such as precision and recall based on corresponding input and output seems to suit poorly for sequence-to-sequence prediction task, since the objective is not to predict an exact match, but to evaluate the amount of text overlap, grammatical quality and meaning of the generated text. Therefore, Bilingual Evaluation Understudy [BLEU] score (Papineni et al., 2002) was used for quality evaluation, which correlates better to human judgement. BLEU score essentially indicates similarity between a predicted and a reference sequence. While BLEU measures precision of the sequence as an overlap, there is another metric called Recall Oriented Understudy [ROUGE] (Lin et al., 2004), which is complimentary to the BLEU score.

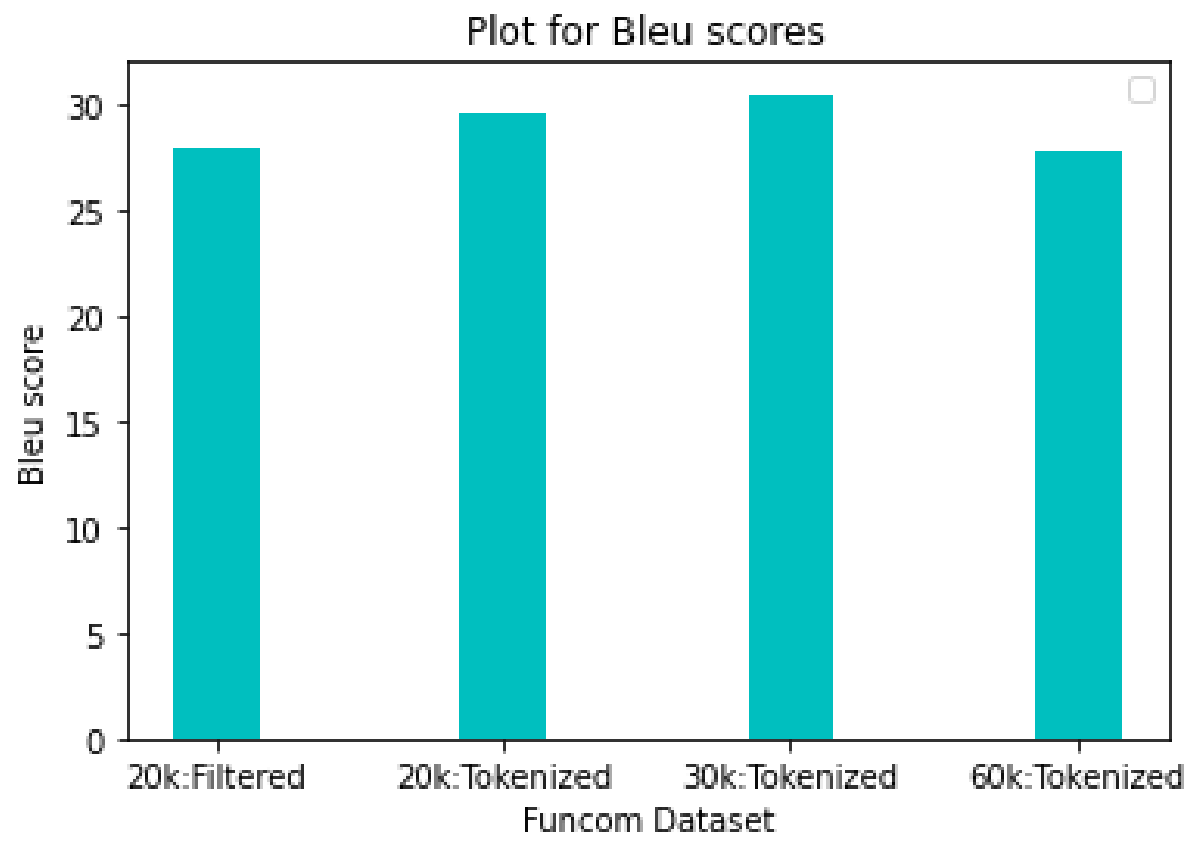

| | rouge-2 | rouge-1 | rouge-3 | rouge-4 | rouge-l | rouge-w |
|---|---|---|---|---|---|---|
| **f** | 0.1990 | 0.3920 | 0.1177 | 0.0865 | 0.4314 | 0.2862 |
| **p** | 0.2215 | 0.4352 | 0.1284 | 0.0927 | 0.4686 | 0.3971 |
| **r** | 0.1972 | 0.3875 | 0.1171 | 0.0848 | 0.4251 | 0.2440 |

Rouge scores based on individual samples [Funcom:60k]

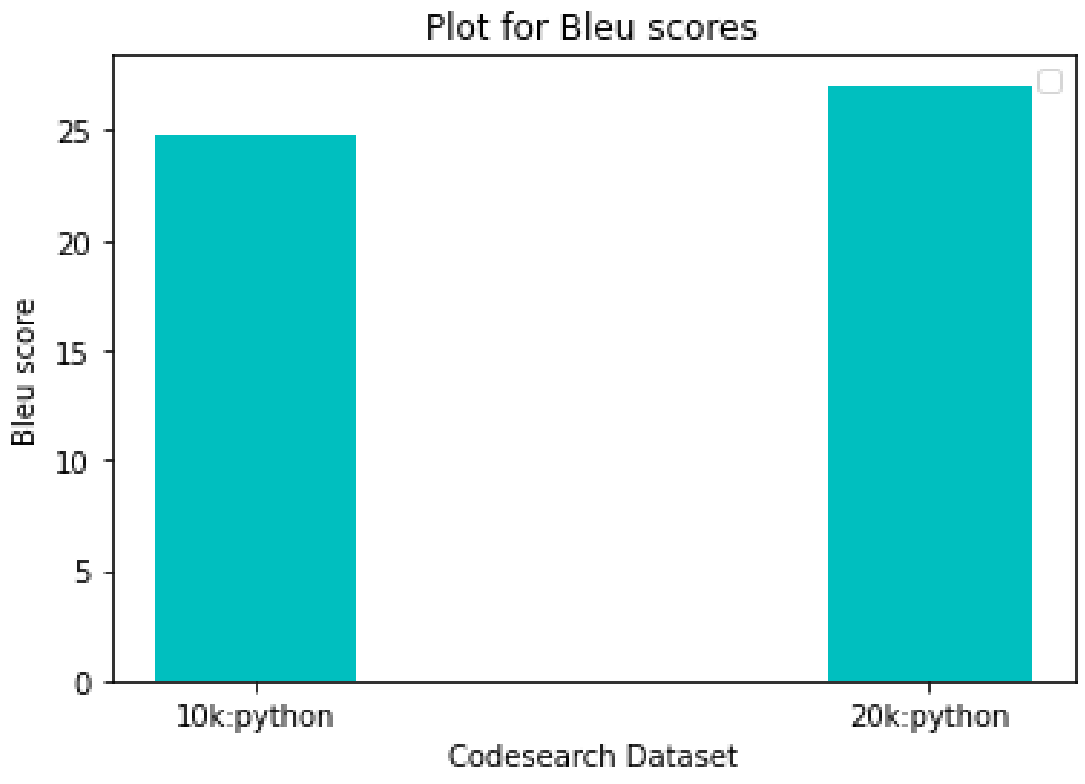

| | rouge-4 | rouge-1 | rouge-3 | rouge-2 | rouge-l | rouge-w |
|---|---|---|---|---|---|---|
| **f** | 0.2824 | 0.5519 | 0.3352 | 0.4111 | 0.5833 | 0.3600 |
| **p** | 0.3037 | 0.5982 | 0.3604 | 0.4423 | 0.6221 | 0.5301 |
| **r** | 0.2804 | 0.5476 | 0.3318 | 0.4071 | 0.5776 | 0.2930 |

Rouge scores based on individual samples [Codesearch:20k]

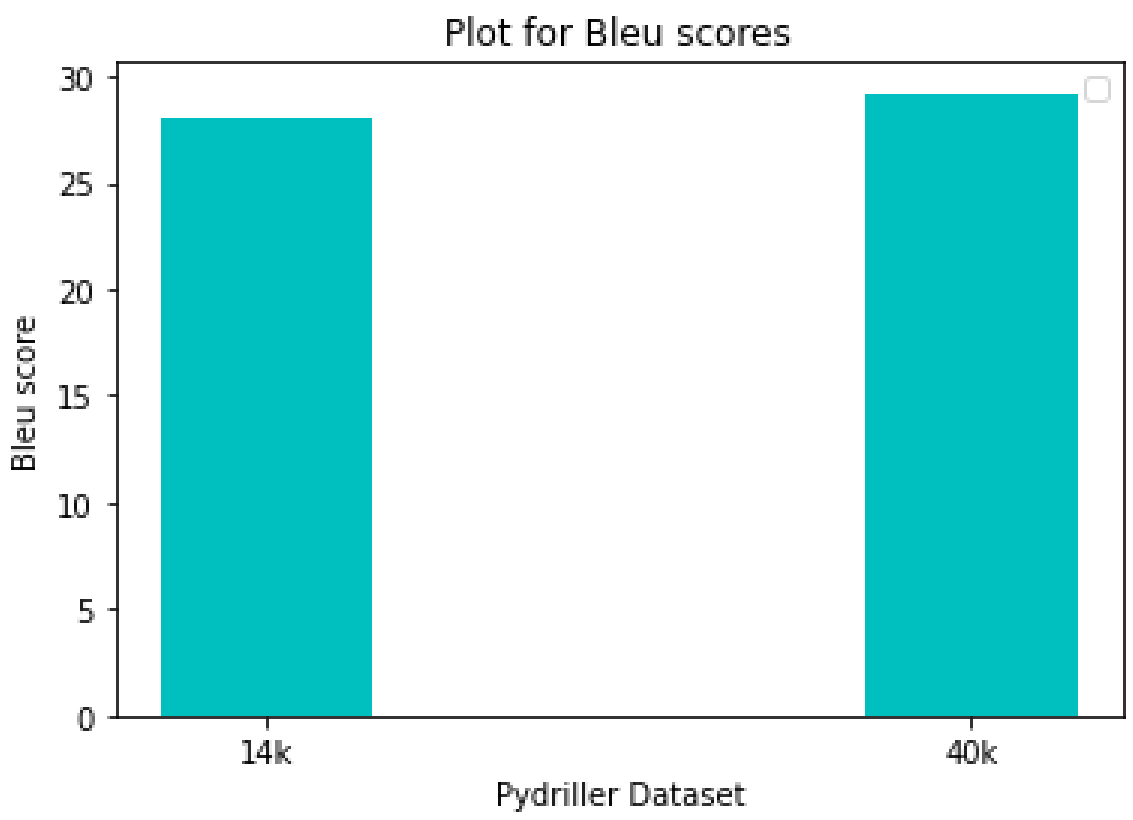

| | rouge-4 | rouge-1 | rouge-2 | rouge-3 | rouge-l | rouge-w |
|---|---|---|---|---|---|---|
| **f** | 0.0112 | 0.1905 | 0.0640 | 0.0177 | 0.2276 | 0.1176 |
| **p** | 0.0115 | 0.2254 | 0.0741 | 0.0190 | 0.2584 | 0.1873 |
| **r** | 0.0120 | 0.1911 | 0.0655 | 0.0185 | 0.2269 | 0.0999 |

Rouge scores based on individual samples [Pydriller:40k]

| Original Comment | Predicted Comment |
|---|---|
| Gets a value object by its key | Returns the value of the specified attribute |
| Returns a string representation without causing additional API calls | Returns a string representation of this object |
| Shows about box on the screen | Shows the dialog |

Comparison of outputs

It can be observed from the predicted output that the model generated a generic summary in several cases, which is ideal for generalization but also results in loss of context. It is also important to note that higher scores on standardized datasets can often leave the problem to be under-constrained. Human annotated summaries that are constrained in nature can have more important terms given a sentence. This annotated data is a better representation of the developer's understanding of a code segment, since factors such as relevance, fluency, consistency, and coherence can be considered. Moreover, with human evaluation, methods such as factual inconsistency and fluency issues can be used to manually review the model results and if inconsistencies are found, it can either be possible to rationalize them or create exceptions and feeding them back to model with corrections. The model in terms of diversification of outputs can be more precisely evaluated with human judgement. Thus, weak reflection of human judgment in the scores can undermine the importance of the results, which is why involvement of manual annotation and evaluation plays a strong role in the development of data-driven summarization techniques.

## Conclusion

In this project, source code summarization is defined as a Natural Language Processing problem to be solved with sequence-to-sequence attention-based models. A custom dataset was generated by mining popular GitHub repositories and annotating the commit message as comment and commit source code difference as the code segment. The datasets were extracted, explored, and applied to train the models generating results that were evaluated upon. Source code belonged to programming languages such as C/C++, Python, Java, and in the future experiments. The generated summaries were evaluated with BLEU and ROUGE scores and the importance of human evaluation was discussed. The future work will focus on understanding convoluted code segments in C and applying beam search to decode the sequences. Furthermore, generative models may be experimented with to branch out the project with program induction techniques.